\documentclass[preprint,showpacs,preprintnumbers,amsmath,amssymb]{revtex4}

\usepackage{graphicx}
\usepackage{dcolumn}
\usepackage{bm}
\usepackage{epsfig}

\newcommand{\mylab}[1]{\label{#1}}
\graphicspath{{.}}

\begin{document}

\title{Drops on an inclined heterogeneous substrate: onset of sliding motion}

\author{Uwe Thiele} 

\affiliation{Max-Planck-Institut f\"ur Physik komplexer Systeme, N{\"o}thnitzer Str.\ 38, 
D-01187 Dresden, Germany}

\author{Edgar Knobloch}

\affiliation{Department of Physics, University of
  California, Berkeley, CA 94720, USA}

\date{\today}

\begin{abstract}
Pinning and depinning of drops on an inclined heterogeneous substrate is studied
as a function of the inclination and heterogeneity amplitude. Two types of heterogeneity 
are considered: a hydrophobic defect that blocks the droplet in front, and a hydrophilic 
one that holds it at the back. Two different types of depinning leading to sliding motion
are identified, and the resulting stick-slip motion is studied numerically.
\end{abstract}


\maketitle

It is well known that liquid drops on an ideally smooth substrate move in response 
to external gradients. For example, a drop on an inclined substrate slides downslope 
in response to the gradient of potential energy \cite{PFL01,Thie01}. Likewise a 
droplet in a temperature gradient will move towards higher temperatures as a result 
of Marangoni forces caused by surface tension gradients \cite{Broc89}. Alternatively, 
a wettability gradient induced by a chemical grading of the substrate also causes 
droplet motion. In order to minimize its energy the droplet will move towards the 
most wettable region \cite{Raph88,ChWh92}. Although on ideally smooth substrates 
droplets will move even for arbitrarily small gradients, this is not the case for 
the 'real' substrates used in experiments. There the onset of contact line motion
is strongly influenced by chemical or physical heterogeneities of the substrate 
%
%
and a finite driving force is necessary to overcome the pinning influence of the 
heterogeneities \cite{ScGa85,NaGa94,Marm96,QAD98,ScWo00,RoFo01,TBBB03}.
On the smaller, atomic scale
surface heterogeneities can trap droplets even on very smooth surfaces.
Indeed, heterogeneities occurring on a micro- or mesoscale are known to affect the 
macroscopic movement of droplets and are responsible, for instance, for 
the observed hysteresis between advancing and receding contact angles 
\cite{Duss79,deGe85,LeJo92}.

The simplest example of depinning is described by the Adler equation \cite{Adle46}
\begin{equation}
\dot\theta=\mu-\sin\theta,
\mylab{adler}
\end{equation}
where $\theta$ represents the position of the droplet, and $\mu>0$ represents the applied
force. When $\mu<1$ this equation has a pair of fixed points, one of which is stable 
and the other unstable. At $\mu=1$ these fixed points annihilate in a saddle-node
bifurcation, but unlike the standard saddle-node bifurcation this bifurcation produces
periodic motion for $\mu>1$. This result is simplest to understand if we write
Eq.~(\ref{adler}) as $\dot\theta=-dV/d\theta$, $V\equiv -\mu \theta-\cos\theta$. Evidently,
Eq.~(\ref{adler}) represents an overdamped particle in a cosinusoidal potential that is progressively
tilted as $\mu$ increases. A 'particle' in a stable equilibrium at a local minimum of
this potential 'spills out' once the tilt becomes large enough that its position
no longer corresponds to a minimum. This occurs precisely at $\mu=1$. The periodic
motion present for $\mu>1$ corresponds to the particle sliding down the resulting
'washboard' potential. The period of this motion diverges as $(\mu-1)^{-1/2}$ \cite{Stro94}.
The resulting bifurcation is sometimes called a Saddle-Node Infinite PERiod bifurcation 
or 'sniper' for short.

In this paper we explore the process of pinning and depinning of driven droplets on a 
heterogeneous substrate. For simplicity we consider the case of gravitational forcing 
on an inclined substrate with a heterogeneous disjoining potential with a well-defined 
spatial period such as might arise from spatially varying wetting properties resulting 
from chemical heterogeneity. This formulation avoids complications arising from changes 
in surface elevation of the substrate (surface roughness) while retaining the essence 
of the pinning phenomenon. In addition we focus on nanoscale droplets for which we can
solve the governing equation for both the droplet profile and the precursor film,
without involving the matched asymptotic expansions required for larger droplets. We
avoid energy methods since these do not permit us to study time-dependent phenomena.

A two-dimensional liquid droplet on an inhomogeneous solid substrate subject
to a horizontal force $\mu$ (Fig.\ref{sketch2}) is described by an evolution equation 
for the film thickness profile $h(x,t)$ derived from the Navier-Stokes equation
using the long-wave approximation \cite{ODB97}:
\begin{equation}
\partial_t\,h\,=\,-\partial_x\, \left\{\frac{h^3}{3\eta}\,\left[ 
\partial_x\,(\gamma\partial_{xx} h + \Pi(h,x))  + \mu\right]\right\}.
\mylab{film}
\end{equation}
Here $\gamma$ is the surface tension, $\eta$ is the dynamic viscosity,
while $\Pi(h,x)$ is the disjoining pressure that accounts for the wetting 
properties of the heterogeneous substrate \cite{Isra92}.
We use the form $\Pi(h,x)=2S_a d_0^2/h^3 + (S_p(x)/l)\,\exp[(d_0-h)/l]$ 
\cite{Shar93,TNPV02}, where $S_a$ and $S_p$ are the apolar and polar components
of the total spreading coefficient $S=S_a+S_p$, $d_0=0.158$\,nm is the Born 
repulsion length and $l$ is a correlation length \cite{Shar93}, and choose 
$S_a>0$ and $S_p<0$, thereby combining a stabilizing long-range van der Waals 
liquid-solid interaction with a destabilizing short-range polar interaction. 
The latter contains the influence of surface coating and wettability defects,
and crucially influences the static contact angle \cite{Shar93}. When $\mu=0$ 
the resulting model describes static droplets with a finite mesoscopic
equilibrium contact angle sitting on an ultrathin precursor film. However,
any qualitatively similar disjoining pressure yields like results, as 
shown for dewetting in \cite{TNPV02} and for chemically driven running droplets 
in \cite{JBT05}.


We nondimensionalize Eq.~(\ref{film}) using the scales 
$3\eta \gamma/\kappa^2 l$ for time, $l$ for the film thickness and 
$\sqrt{l\gamma/\kappa}$ for the lateral coordinate, where $\kappa=(|S_p|/l)\,\exp(d_0/l)$.
In addition we define the dimensionless quantities $b=(2S_{a}d_0^2/|S_p|l^2)\,\exp(-d_0/l)$
and $\alpha=(\gamma l/\kappa^3)^{1/2}(\mu/\rho)$.
Thus for gravitational forcing $\alpha$ measures the inclination
of the substrate, and we refer to it as the inclination. The loading of
the system (relevant for gravitational forcing) is measured by the mean
film height ${\bar h}\equiv L^{-1}\int_0^L\, h(x)\,dx$.


Figures \ref{profang}(a,b) show sample steady state profiles for two cases:
(a) a hydrophilic defect, (b) a hydrophobic defect. Both are described by
\begin{equation}
\Pi(h,x)\,=\,\frac{b}{h^3} - [1 + \epsilon\xi(x)]\,e^{-h},
\end{equation}
with
\begin{equation}
\xi(x)\,=\,\{2\,\mbox{cn}[2K(k)x/L,k]\}^2 - \Delta,
\mylab{kappa3}
\end{equation}
where $K(k)$ is the complete elliptic integral of the first kind and $\Delta$ 
is such that the average of $\xi(x)$ over a spatial period vanishes. We use the 
logarithmic measure $s\equiv-\log(1-k)$ to quantify the steepness of the heterogeneity 
profiles (Fig.~\ref{profang}(a,b)). These correspond to hydrophilic ($\epsilon<0$) or
hydrophobic ($\epsilon>0$) defects.
In (a) the droplet is held at the back by a hydrophilic defect and develops a 
prominent shoulder as $\alpha$ increases just prior to depinning. In contrast in (b) the
hydrophobic defect blocks the droplet and its profile steepens with increasing
$\alpha$. The profiles at depinning are shaded. Figures \ref{profang}(c,d) 
show the advancing and receding mesoscopic contact angles, measured at the inflection points
of the drop profile, as a function of $\alpha$. 
For a droplet pinned at the back (Fig.~\ref{profang}(c)) the advancing 
[receding] angle decreases [increases] for small but increasing inclination $\alpha$. However, 
once the droplet starts developing a shoulder at the back the receding angle decreases again 
until depinning occurs. The situation differs for a droplet pinned at the front (Fig.~\ref{profang}(d)).
In this case both angles increase with $\alpha$ but drop just prior to depinning (Fig.\,\ref{profang}).

The depinning process corresponds to the loss of stability of the pinned drop. The stability
calculation \cite{ThKn06} reveals two mechanisms that lead to depinning. The first is via
a sniper bifurcation (i.e., a steady state bifurcation) and prevails for hydrophobic defects
with small wettability contrast \cite{ThKn06} and for hydrophilic defects. Figure \ref{bifphilall}(a) 
shows a typical bifurcation diagram for the latter case as a function of increasing $\alpha$. 
The figure shows the $L^2$ norm of $\delta h\equiv h(x)-{\bar h}$ for pinned drops 
and its time-average after depinning. Although there are two saddle-node bifurcations in the diagram
time integration (open circles) shows that the upper part of the branch of pinned drops is stable
until the rightmost saddle-node bifurcation. Thereafter the solutions are time-dependent but periodic
(open triangles). The inset shows that near the saddle-node the period diverges like $(\alpha-\alpha_c)^{-1/2}$
and hence that in this case depinning corresponds to a sniper bifurcation. Figures \ref{bifphilall}(b,c)
show space-time plots of the resulting motion for (b) $\alpha\gtrsim\alpha_c$, and (c) $\alpha=0.04$.
In (b) the drop spends a long time in a nearly stationary state while slowly spreading downstream,
before it abruptly breaks off and moves towards the next defect. In contrast in (c) the drop flows 
more or less at constant speed downslope, although the location of the defect remains visible in the 
space-time plot.



Figure \ref{foldalep}(a) shows the location of the two saddle-nodes in the $(\epsilon,\alpha)$ plane.
In the case of a hydrophilic defect ($\epsilon<0$) the saddle-nodes are always present; the one at 
larger $\alpha$ corresponds to the depinning bifurcation. For fixed ${\bar h}$ and large $L$ the 
critical $\alpha$ decreases as $1/L$ (not shown), as expected on the basis of simple loading ideas.
However, the figure also shows that something else happens for sufficiently hydrophobic defects. 
Here the saddle-nodes annihilate at $\epsilon\approx0.6$, and depinning now occurs via a Hopf 
bifurcation (dashed line). The resulting bifurcation diagram (Fig.~\ref{bifhopfall}(a)) shows 
that the range of stable pinned profiles overlaps with the range of periodic states generated 
by the instability. Thus in this case the branch of periodic solutions loses 
stability at a saddle-node bifurcation as $\alpha$ decreases, and the system settles into a 
steady pinned state in a hysteretic transition. Figures \ref{bifhopfall}(b,c) show space-time 
plots of the periodic state near this transition and further away. Here the depinning 
is as abrupt as in Fig.~\ref{bifphilall}(b) but without the slow downstream leakage seen in 
the latter figure. The resulting dynamics strongly resemble stick-slip motion. However, further 
away from the transition the depinned states in both cases look quite alike: in both cases the 
droplet travels at almost constant speed, only slightly modulated by the heterogeneity.


The advancing and receding angles at depinning (shaded profiles in Fig.~\ref{profang})
shown in Fig.\,\ref{foldalep}(b) provide a measure of the contact angle hysteresis observed 
macroscopically. In the case of a hydrophobic defect at the front ($\epsilon>0$) both angles 
increase nearly linearly with defect strength, 
and continue to do so even for oscillatory depinning ($\epsilon\gtrsim 0.6$); 
the small hook visible in the figure near this transition indicates that the Hopf 
bifurcation sets in prior to the disappearance of the saddle-node bifurcations. The behavior is more 
intricate when the pinning is by a hydrophilic defect at the back ($\epsilon<0$). In this case the 
role of the two angles is reversed, and both decrease nearly linearly with slopes identical to those 
in the $\epsilon>0$ case. For $\epsilon<-0.2$, however, the receding angle reverses tendency and 
starts to increase again, while the advancing angle continues to decrease. This change in behavior
is a consequence of the stretching of the drop with increasing inclination just prior to depinning:
for $\epsilon\lesssim-0.2$ gravity drags the main body of liquid downstream (to the right) but the 
spot of higher wettability traps part of it upstream. For fixed $\alpha$ the latter effect becomes 
more pronounced as $|\epsilon|$ increases, cf.~Fig.\,\ref{foldalep}(b).


We have examined two types of pinning: pinning by a hydrophilic defect at the
back of the droplet, and pinning by a hydrophobic defect in front of it, and
identified two mechanisms whereby pinning takes place. In the case of a sufficiently 
large hydrophilic defect the droplet stretches markedly just prior to depinning as
the substrate inclination increases; the inclined droplet loses stability 
at a saddle-node bifurcation, resulting in periodic motion as the
droplet slides over a periodic array of hydrophilic defects. We have referred
to this type of bifurcation as the 'sniper'. The periodic motion that results
is slow when the droplet is stretching, and fast once the droplet breaks away
from a defect and spills onto the next one. The situation is richer for
hydrophobic defects that pin the droplet by blocking it. In this case in addition
to the steady state sniper bifurcation a new depinning mechanism was observed:
the droplet loses stability to an oscillatory mode prior to depinning. A mode
of this type cannot be identified by standard energy arguments. In the example shown 
this bifurcation is hysteretic. The two depinning scenarios are distinguished primarily 
by the average speed of the droplet near the depinning transition. In the sniper 
scenario this speed vanishes as $(\alpha-\alpha_c)^{1/2}$; in the latter it is 
finite. At larger values of $\alpha$ both scenarios lead to broadly similar 
dynamics: more-or-less uniform sliding motion modulated by passage over defects.
It is noteworthy that no Hopf bifurcation occurs when the wettability profile
is sinusoidal \cite{ThKn06}.

Many depinning phenomena in physics may be understood using the sniper scenario.
Usually this is so in systems with a continuous symmetry such as invariance under
translations. In the absence of a heterogeneity spatially periodic structures
may undergo a spontaneous parity-breaking bifurcation that breaks the left-right symmetry
of the pattern and produces a drift. The direction of the drift is then determined by
the associated tilt of the structure \cite{CGG89}. In this case the drift speed of the
structure vanishes as the square root of the distance from the parity-breaking 
bifurcation. However, in the presence of spatial heterogeneities the
situation changes dramatically because near the bifurcation even small amplitude 
heterogeneities suffice to pin the tilted structure. A detailed study of this
regime \cite{DHK97} shows that while some depinning events are indeed analogous
to the behavior described by the Adler equation, a quite different depinning
mechanism is present as well. Here the tilted state first undergoes
a Hopf bifurcation that produces back-and-forth rocking motion of the tilted
structure, but no net translation. As a parameter increases the amplitude 
of this oscillation increases leading to a global bifurcation involving an unstable 
fixed point and its translate by one period. This bifurcation generates oscillations 
with a nonzero mean drift, and this net drift increases with further increase in 
the parameter. The present system differs in the absence of left-right symmetry,
but a global bifurcation that changes the topology of the limit cycle produced
in the Hopf bifurcation from a libration to a rotation must still take place.
Such a bifurcation can occur if the Hopf bifurcation is in fact supercritical.
Consistency with the Fig.~\ref{bifhopfall}(a) requires that the branch of 
periodic states must go through a pair of saddle-node bifurcations
to produce stable states of the type shown in Fig.~\ref{bifhopfall}(b).

This work was supported by NASA, NSF and EU under grants NNC04GA47G (EK,UT), 
DMS-0305968 (EK) and MRTN-CT-2004-005728 (UT).


\begin{thebibliography}{25}
\expandafter\ifx\csname natexlab\endcsname\relax\def\natexlab#1{#1}\fi
\expandafter\ifx\csname bibnamefont\endcsname\relax
  \def\bibnamefont#1{#1}\fi
\expandafter\ifx\csname bibfnamefont\endcsname\relax
  \def\bibfnamefont#1{#1}\fi
\expandafter\ifx\csname citenamefont\endcsname\relax
  \def\citenamefont#1{#1}\fi
\expandafter\ifx\csname url\endcsname\relax
  \def\url#1{\texttt{#1}}\fi
\expandafter\ifx\csname urlprefix\endcsname\relax\def\urlprefix{URL }\fi
\providecommand{\bibinfo}[2]{#2}
\providecommand{\eprint}[2][]{\url{#2}}

\bibitem[{\citenamefont{Podgorski et~al.}(2001)\citenamefont{Podgorski,
  Flesselles, and Limat}}]{PFL01}
\bibinfo{author}{\bibfnamefont{T.}~\bibnamefont{Podgorski}},
  \bibinfo{author}{\bibfnamefont{J.-M.} \bibnamefont{Flesselles}},
  \bibnamefont{and} \bibinfo{author}{\bibfnamefont{L.}~\bibnamefont{Limat}},
  \bibinfo{journal}{Phys. Rev. Lett.} \textbf{\bibinfo{volume}{87}},
  \bibinfo{pages}{036102} (\bibinfo{year}{2001}).

\bibitem[{\citenamefont{Thiele et~al.}(2001)\citenamefont{Thiele, Velarde,
  Neuffer, Bestehorn, and Pomeau}}]{Thie01}
\bibinfo{author}{\bibfnamefont{U.}~\bibnamefont{Thiele}},
  \bibinfo{author}{\bibfnamefont{M.~G.} \bibnamefont{Velarde}},
  \bibinfo{author}{\bibfnamefont{K.}~\bibnamefont{Neuffer}},
  \bibinfo{author}{\bibfnamefont{M.}~\bibnamefont{Bestehorn}},
  \bibnamefont{and} \bibinfo{author}{\bibfnamefont{Y.}~\bibnamefont{Pomeau}},
  \bibinfo{journal}{Phys. Rev. E} \textbf{\bibinfo{volume}{64}},
  \bibinfo{pages}{061601} (\bibinfo{year}{2001}).

\bibitem[{\citenamefont{Brochard}(1989)}]{Broc89}
\bibinfo{author}{\bibfnamefont{F.}~\bibnamefont{Brochard}},
  \bibinfo{journal}{Langmuir} \textbf{\bibinfo{volume}{5}},
  \bibinfo{pages}{432} (\bibinfo{year}{1989}).

\bibitem[{\citenamefont{Rapha{\"e}l}(1988)}]{Raph88}
\bibinfo{author}{\bibfnamefont{E.}~\bibnamefont{Rapha{\"e}l}},
  \bibinfo{journal}{C. R. Acad. Sci. Ser. II} \textbf{\bibinfo{volume}{306}},
  \bibinfo{pages}{751} (\bibinfo{year}{1988}).

\bibitem[{\citenamefont{Chaudhury and Whitesides}(1992)}]{ChWh92}
\bibinfo{author}{\bibfnamefont{M.~K.} \bibnamefont{Chaudhury}}
  \bibnamefont{and} \bibinfo{author}{\bibfnamefont{G.~M.}
  \bibnamefont{Whitesides}}, \bibinfo{journal}{Science}
  \textbf{\bibinfo{volume}{256}}, \bibinfo{pages}{1539} (\bibinfo{year}{1992}).

\bibitem[{\citenamefont{Schwartz and Garoff}(1985)}]{ScGa85}
\bibinfo{author}{\bibfnamefont{L.~W.} \bibnamefont{Schwartz}} \bibnamefont{and}
  \bibinfo{author}{\bibfnamefont{S.}~\bibnamefont{Garoff}},
  \bibinfo{journal}{Langmuir} \textbf{\bibinfo{volume}{1}},
  \bibinfo{pages}{219} (\bibinfo{year}{1985}).

\bibitem[{\citenamefont{Nadkarni and Garoff}(1994)}]{NaGa94}
\bibinfo{author}{\bibfnamefont{G.~D.} \bibnamefont{Nadkarni}} \bibnamefont{and}
  \bibinfo{author}{\bibfnamefont{S.}~\bibnamefont{Garoff}},
  \bibinfo{journal}{Langmuir} \textbf{\bibinfo{volume}{10}},
  \bibinfo{pages}{1618} (\bibinfo{year}{1994}).

\bibitem[{\citenamefont{Marmur}(1996)}]{Marm96}
\bibinfo{author}{\bibfnamefont{A.}~\bibnamefont{Marmur}},
  \bibinfo{journal}{Colloid Surf. A-Physicochem. Eng. Asp.}
  \textbf{\bibinfo{volume}{116}}, \bibinfo{pages}{55} (\bibinfo{year}{1996}).

\bibitem[{\citenamefont{Qu{\'e}r{\'e} et~al.}(1998)\citenamefont{Qu{\'e}r{\'e},
  Azzopardi, and Delattre}}]{QAD98}
\bibinfo{author}{\bibfnamefont{D.}~\bibnamefont{Qu{\'e}r{\'e}}},
  \bibinfo{author}{\bibfnamefont{M.~J.} \bibnamefont{Azzopardi}},
  \bibnamefont{and} \bibinfo{author}{\bibfnamefont{L.}~\bibnamefont{Delattre}},
  \bibinfo{journal}{Langmuir} \textbf{\bibinfo{volume}{14}},
  \bibinfo{pages}{2213} (\bibinfo{year}{1998}).

\bibitem[{\citenamefont{Sch{\"a}ffer and Wong}(2000)}]{ScWo00}
\bibinfo{author}{\bibfnamefont{E.}~\bibnamefont{Sch{\"a}ffer}}
  \bibnamefont{and} \bibinfo{author}{\bibfnamefont{P.~Z.} \bibnamefont{Wong}},
  \bibinfo{journal}{Phys. Rev. E} \textbf{\bibinfo{volume}{61}},
  \bibinfo{pages}{5257} (\bibinfo{year}{2000}).

\bibitem[{\citenamefont{Roura and Fort}(2001)}]{RoFo01}
\bibinfo{author}{\bibfnamefont{P.}~\bibnamefont{Roura}} \bibnamefont{and}
  \bibinfo{author}{\bibfnamefont{J.}~\bibnamefont{Fort}},
  \bibinfo{journal}{Phys. Rev. E} \textbf{\bibinfo{volume}{64}},
  \bibinfo{pages}{011601} (\bibinfo{year}{2001}).

\bibitem[{\citenamefont{Thiele et~al.}(2003)\citenamefont{Thiele, Brusch,
  Bestehorn, and B{\"a}r}}]{TBBB03}
\bibinfo{author}{\bibfnamefont{U.}~\bibnamefont{Thiele}},
  \bibinfo{author}{\bibfnamefont{L.}~\bibnamefont{Brusch}},
  \bibinfo{author}{\bibfnamefont{M.}~\bibnamefont{Bestehorn}},
  \bibnamefont{and} \bibinfo{author}{\bibfnamefont{M.}~\bibnamefont{B{\"a}r}},
  \bibinfo{journal}{Eur. Phys. J. E} \textbf{\bibinfo{volume}{11}},
  \bibinfo{pages}{255} (\bibinfo{year}{2003}).

\bibitem[{\citenamefont{Dussan}(1979)}]{Duss79}
\bibinfo{author}{\bibfnamefont{E.~B.} \bibnamefont{Dussan}},
  \bibinfo{journal}{Ann. Rev. Fluid Mech.} \textbf{\bibinfo{volume}{11}},
  \bibinfo{pages}{371} (\bibinfo{year}{1979}).

\bibitem[{\citenamefont{de~Gennes}(1985)}]{deGe85}
\bibinfo{author}{\bibfnamefont{P.-G.} \bibnamefont{de~Gennes}},
  \bibinfo{journal}{Rev. Mod. Phys.} \textbf{\bibinfo{volume}{57}},
  \bibinfo{pages}{827} (\bibinfo{year}{1985}).

\bibitem[{\citenamefont{Leger and Joanny}(1992)}]{LeJo92}
\bibinfo{author}{\bibfnamefont{L.}~\bibnamefont{Leger}} \bibnamefont{and}
  \bibinfo{author}{\bibfnamefont{J.~F.} \bibnamefont{Joanny}},
  \bibinfo{journal}{Rep. Prog. Phys.} \textbf{\bibinfo{volume}{55}},
  \bibinfo{pages}{431} (\bibinfo{year}{1992}).

\bibitem[{\citenamefont{Adler}(1946)}]{Adle46}
\bibinfo{author}{\bibfnamefont{R.}~\bibnamefont{Adler}},
  \bibinfo{journal}{Proc. I.R.E. Waves Electrons}
  \textbf{\bibinfo{volume}{34}}, \bibinfo{pages}{351} (\bibinfo{year}{1946}).

\bibitem[{\citenamefont{Strogatz}(1994)}]{Stro94}
\bibinfo{author}{\bibfnamefont{S.~H.} \bibnamefont{Strogatz}},
  \emph{\bibinfo{title}{Nonlinear Dynamics and Chaos}}
  (\bibinfo{publisher}{Addison-Wesley}, \bibinfo{year}{1994}).

\bibitem[{\citenamefont{Oron et~al.}(1997)\citenamefont{Oron, Davis, and
  Bankoff}}]{ODB97}
\bibinfo{author}{\bibfnamefont{A.}~\bibnamefont{Oron}},
  \bibinfo{author}{\bibfnamefont{S.~H.} \bibnamefont{Davis}}, \bibnamefont{and}
  \bibinfo{author}{\bibfnamefont{S.~G.} \bibnamefont{Bankoff}},
  \bibinfo{journal}{Rev. Mod. Phys.} \textbf{\bibinfo{volume}{69}},
  \bibinfo{pages}{931} (\bibinfo{year}{1997}).

\bibitem[{\citenamefont{Israelachvili}(1992)}]{Isra92}
\bibinfo{author}{\bibfnamefont{J.~N.} \bibnamefont{Israelachvili}},
  \emph{\bibinfo{title}{Intermolecular and Surface Forces}}
  (\bibinfo{publisher}{Academic Press}, \bibinfo{address}{London},
  \bibinfo{year}{1992}).

\bibitem[{\citenamefont{Sharma}(1993)}]{Shar93}
\bibinfo{author}{\bibfnamefont{A.}~\bibnamefont{Sharma}},
  \bibinfo{journal}{Langmuir} \textbf{\bibinfo{volume}{9}},
  \bibinfo{pages}{861} (\bibinfo{year}{1993}).

\bibitem[{\citenamefont{Thiele et~al.}(2002)\citenamefont{Thiele, Neuffer,
  Pomeau, and Velarde}}]{TNPV02}
\bibinfo{author}{\bibfnamefont{U.}~\bibnamefont{Thiele}},
  \bibinfo{author}{\bibfnamefont{K.}~\bibnamefont{Neuffer}},
  \bibinfo{author}{\bibfnamefont{Y.}~\bibnamefont{Pomeau}}, \bibnamefont{and}
  \bibinfo{author}{\bibfnamefont{M.~G.} \bibnamefont{Velarde}},
  \bibinfo{journal}{Colloid Surf. A} \textbf{\bibinfo{volume}{206}},
  \bibinfo{pages}{135} (\bibinfo{year}{2002}).

\bibitem[{\citenamefont{John et~al.}(2005)\citenamefont{John, B{\"a}r, and
  Thiele}}]{JBT05}
\bibinfo{author}{\bibfnamefont{K.}~\bibnamefont{John}},
  \bibinfo{author}{\bibfnamefont{M.}~\bibnamefont{B{\"a}r}}, \bibnamefont{and}
  \bibinfo{author}{\bibfnamefont{U.}~\bibnamefont{Thiele}},
  \bibinfo{journal}{Eur. Phys. J. E} \textbf{\bibinfo{volume}{18}},
  \bibinfo{pages}{183} (\bibinfo{year}{2005}).

\bibitem[{\citenamefont{Thiele and Knobloch}(2006)}]{ThKn06}
\bibinfo{author}{\bibfnamefont{U.}~\bibnamefont{Thiele}} \bibnamefont{and}
  \bibinfo{author}{\bibfnamefont{E.}~\bibnamefont{Knobloch}},
  \bibinfo{journal}{preprint}  (\bibinfo{year}{2006}).

\bibitem[{\citenamefont{Coullet et~al.}(1989)\citenamefont{Coullet, Goldstein,
  and Gunaratne}}]{CGG89}
\bibinfo{author}{\bibfnamefont{P.}~\bibnamefont{Coullet}},
  \bibinfo{author}{\bibfnamefont{R.~E.} \bibnamefont{Goldstein}},
  \bibnamefont{and} \bibinfo{author}{\bibfnamefont{G.~H.}
  \bibnamefont{Gunaratne}}, \bibinfo{journal}{Phys. Rev. Lett.}
  \textbf{\bibinfo{volume}{63}}, \bibinfo{pages}{1954} (\bibinfo{year}{1989}).

\bibitem[{\citenamefont{Dangelmayr et~al.}(1997)\citenamefont{Dangelmayr,
  Hettel, and Knobloch}}]{DHK97}
\bibinfo{author}{\bibfnamefont{G.}~\bibnamefont{Dangelmayr}},
  \bibinfo{author}{\bibfnamefont{J.}~\bibnamefont{Hettel}}, \bibnamefont{and}
  \bibinfo{author}{\bibfnamefont{E.}~\bibnamefont{Knobloch}},
  \bibinfo{journal}{Nonlinearity} \textbf{\bibinfo{volume}{10}},
  \bibinfo{pages}{1093} (\bibinfo{year}{1997}).

\end{thebibliography}
%

\clearpage

\begin{figure}
\centering
\includegraphics[width=0.6\hsize]{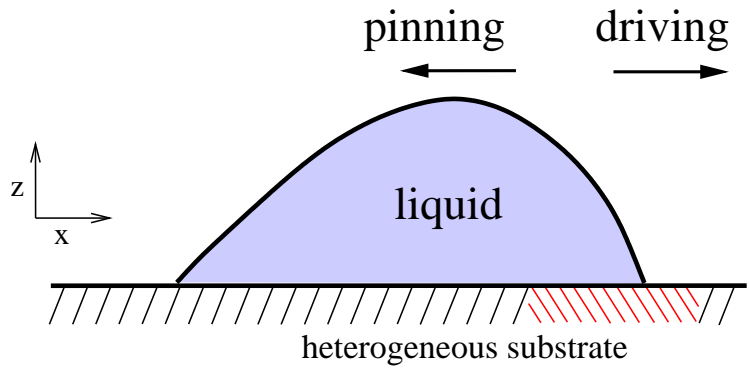}
  \caption{Sketch of a droplet on a heterogeneous substrate subject to
a horizontal force $\mu$ towards the right.
\mylab{sketch2}
}
\end{figure}

\clearpage

\begin{figure}
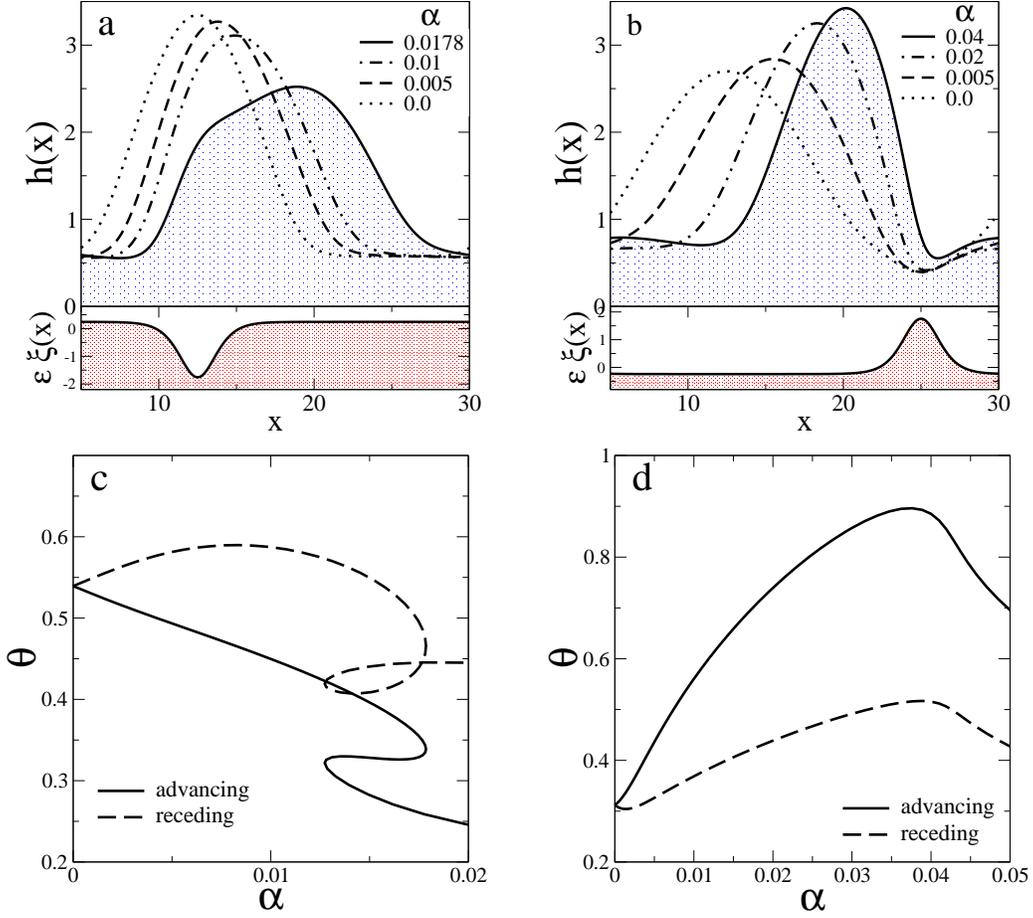

\centering
\includegraphics[width=0.37\hsize]{prof_diffal_b0.1L25h1.5het-1.0s6cos2_square.eps}\hspace{0.05\hsize}
\includegraphics[width=0.37\hsize]{prof_diffal_b0.1L25h1.5het1.0s6cos2_square.eps}\\[1ex]
\includegraphics[width=0.39\hsize]{theta_al_het-1.0b0.1L25h1.5s6cos2_square.eps}\hspace{0.04\hsize}
\includegraphics[width=0.39\hsize]{theta_al_het1.0b0.1L25h1.5s6cos2_square.eps}
  \caption{Characteristics of pinned droplets as a function of the forcing $\alpha$ for localized
hydrophilic [(a) and (c), $\epsilon=-1$] and hydrophobic [(b) and (d), $\epsilon=1$] defects. 
The upper parts of (a) and (b) show steady droplet profiles while the lower parts show the wettability 
profile [Eq.\,(\ref{kappa3}) with $s=6$]. The profile at depinning is shaded. In (a) the droplet is pinned 
by a more wettable defect at the back whereas in (b) it is blocked by a less wettable defect in front.
Panels (c) and (d) show the advancing and receding contact angles $\theta$ 
as a function of $\alpha$. The remaining 
parameters are $L=25$, $b=0.1$, and $\bar{h}=1.5$.
}
\mylab{profang}
\end{figure}

\clearpage

\begin{figure}
\centering
\includegraphics[width=0.9\hsize]{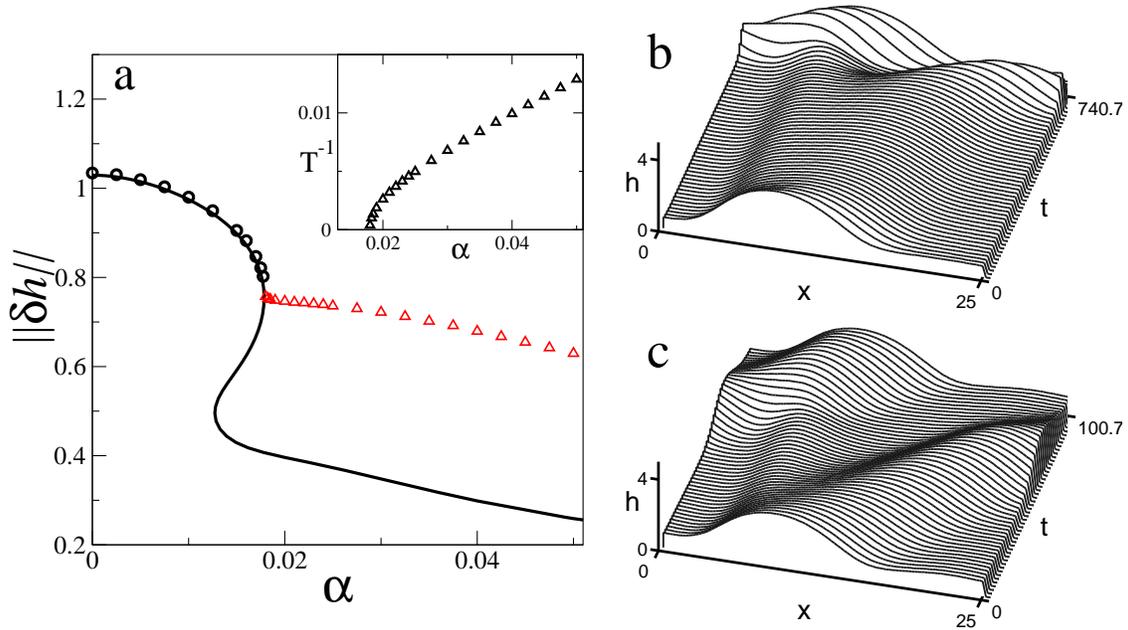}
  \caption{(a) Bifurcation diagram for depinning via a sniper bifurcation for a hydrophilic defect 
[Eq.\,(\ref{kappa3}) with $s=6$] with $\epsilon=-1.0$ and $L=25$, $b=0.1$, $\bar{h}=1.5$. The figure
shows the $L^2$-norm of steady solutions (solid line),
selected steady solutions as obtained by integration in time (circles)
and the time-averaged $L_2$-norm for the unsteady solutions beyond depinning 
(triangles). Inset shows the inverse of the temporal period $T$ for the latter.
The remaining panels show space-time plots over one spatial and temporal period for a 
sliding drop (b) close to depinning at $\alpha=0.0185$ with $T=556.1$, 
and (c) far from depinning at $\alpha=0.04$ with $T=100.7$.
\mylab{bifphilall}
}
\end{figure}

\clearpage

\begin{figure}
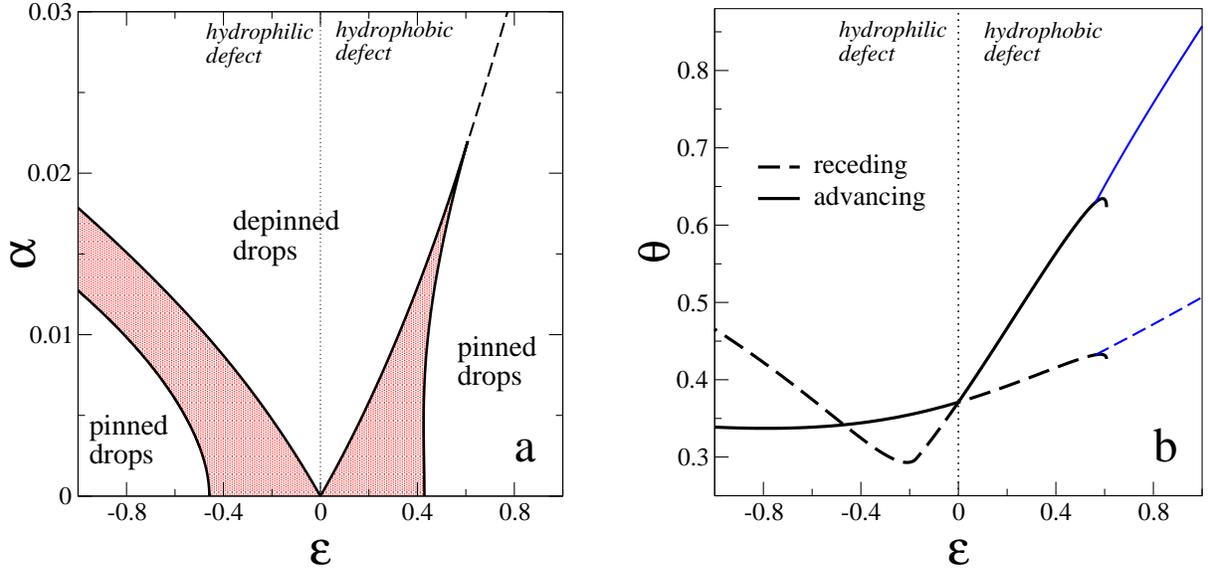

\centering
\includegraphics[width=0.45\hsize]{fold_al_allhet_b0.1L25h1.5s6cos2_inverted.eps}\hspace{0.05\hsize}
\includegraphics[width=0.46\hsize]{theta_fold_al_het_b0.1L25h1.5s6cos2_square.eps}
  \caption{(a) Phase diagram for the depinning transition for localized hydrophilic ($\epsilon<0$) 
or hydrophobic ($\epsilon>0$) defects [Eq.\,(\ref{kappa3}) with $s=6$]. 
The figure focuses on small wettability contrast and $L=25$, $b=0.1$, $\bar{h}=1.5$.
The solid [dashed] lines correspond to saddle-node [Hopf] bifurcations. The latter emerge near 
the cusp at which the two saddle-node bifurcations annihilate for $\epsilon>0$.
(b) Advancing (solid lines) and receding (broken lines) contact angles $\theta$ at the depinning transition
as a function of wettability contrast for a hydrophilic defect at the back ($\epsilon<0$) and hydrophobic 
defect at the front ($\epsilon>0$). Thick [thin] lines refer to depinning through a real [oscillatory] 
mode. 
}
\mylab{foldalep}
\end{figure}

\clearpage

\begin{figure}
\centering
\includegraphics[width=0.9\hsize]{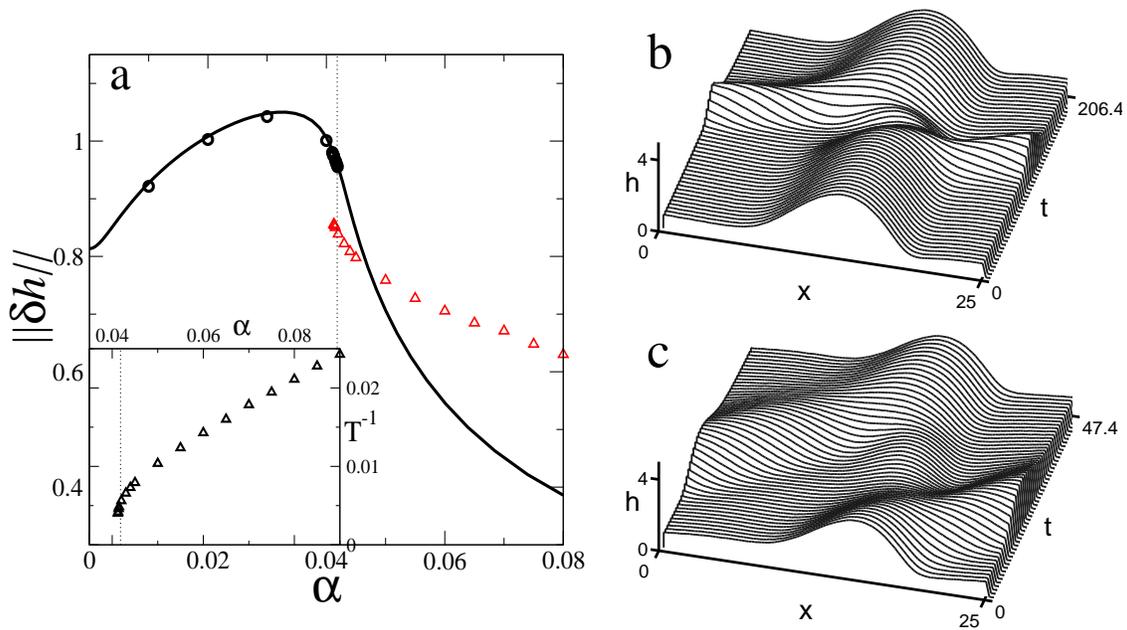}
  \caption{As for Fig.~\ref{bifphilall} but showing depinning via a Hopf bifurcation when $\epsilon=1.0$.
(a) Bifurcation diagram. (b) Space-time plot for $\alpha=0.0415$ with $T=206.4$.
(c) $\alpha=0.08$ with $T=47.4$. The vertical line indicates the location of the Hopf bifurcation as 
obtained from linear stability theory \cite{ThKn06}.
\mylab{bifhopfall}
}
\end{figure}

\end{document}